\documentclass[aps,prl,twocolumn,floatfix,nofootinbib,showpacs]{revtex4}
\usepackage{graphicx}
\usepackage{color}
\begin{document}

\title{Could the Excess Seen at $124-126$ GeV be due to the 
Randall-Sundrum Radion?}
\renewcommand{\thefootnote}{\fnsymbol{footnote}}

\author{
Kingman Cheung$^{1,2}$ and Tzu-Chiang Yuan$^3$}
\affiliation{
$^1$ Department of Physics, National Tsing Hua University,
Hsinchu 300, Taiwan \\
$^2$ Division of Quantum Phases and Devices, School of Physics, 
Konkuk University, Seoul 143-701, Republic of Korea \\
$^3$ Institute of Physics, Academia Sinica, Nangang, Taipei 11529, Taiwan 
}
\renewcommand{\thefootnote}{\arabic{footnote}}
\date{\today}

\begin{abstract}
  Current Higgs boson searches in various channels at the LHC point to an
  excess at around 124--126 GeV due to a possibly standard-model-like
  Higgs boson.  If one examines more closely the channels
  ($\gamma\gamma$, $WW^*$, and $ZZ^*$) that have excess, this ``Higgs
  boson'' may be the Randall-Sundrum radion $\phi$.  Because of the
  trace anomaly the radion has stronger couplings to the photon and
  gluon pairs. Thus, it will enhance the production rates into $gg$
  and $\gamma\gamma$ while those for $WW^*$, $ZZ^*$ and $b\bar b$ are
  reduced relative to their standard-model values.  We show that it can
  match well with the data from CMS for $m_\phi = 124$ GeV and the
  required scale $\Lambda_\phi \sim \langle \phi \rangle$ is about 0.68 TeV.
\end{abstract}
\maketitle

{\it Introduction.}--  
With tremendous speculations before December 13, 2011 the first glimpse 
of the Higgs boson was revealed on that day. Both ATLAS \cite{atlas}
and CMS \cite{cms} saw some excess of events of the Higgs decays in the
$H\to \gamma\gamma$, 
$H\to WW^*  \to \ell \nu \ell \nu$ and $H \to Z Z^* \to 4\ell$ channels.
If one examines more closely these channels, one may notice that
the excessive channels exhibit some correlations, even though it is still too
early to say anything concrete.  According to the CMS data \cite{cms}
for the Higgs mass $m_H = 124$ GeV, the excess relative to the 
corresponding standard model (SM) values are
\begin{eqnarray}
\sigma(H) \times B(H \to b\bar b) / \sigma B_{\rm SM} & \sim  & 
1.1\;^{+1.5}_{-1.6} \nonumber \\
\sigma(H) \times B(H \to \tau\tau) / \sigma B_{\rm SM}  & \sim  & 
0.8\;^{+1.2}_{-1.3} \nonumber \\
\sigma(H) \times B(H \to \gamma\gamma ) / \sigma B_{\rm SM} & \sim  & 
2.1\;^{+0.6}_{-0.7}  \label{exp}\\
\sigma(H) \times B(H \to W W^*)/ \sigma B_{\rm SM} & \sim  & 
0.7\;^{+0.4}_{-0.6}  \nonumber \\
\sigma(H) \times B(H \to ZZ^* \to 4\ell) / \sigma B_{\rm SM} & \sim  & 
0.5\;^{+1.1}_{-0.7}
\nonumber 
\end{eqnarray}
where $\sigma B_{\rm SM}$ denotes the cross section $\sigma(H)$ times 
the corresponding branching ratio for the SM
(ATLAS \cite{atlas} also has the ratio of $\gamma\gamma$ production
rate larger than the SM one.)
At face value, except for the $\gamma\gamma$ channel,
almost all are slightly suppressed relative to the SM cross sections.  
Note that these results consist of large errors.  
If we take these numbers seriously, the 
branching ratios of the $124-126$ GeV ``Higgs boson'' observed have 
to be modified.
One possible way is to add an unobserved channel, e.g., dijet or invisible
particles, such that the Higgs decays into $b\bar b$, $\tau\tau$, $WW^*$,
 and $ZZ^*$
are reduced, while at the same time the $\gamma\gamma$ channel has to be
enhanced by a relatively large amount.
In this work, we point out that the Randall-Sundrum (RS) radion, with 
enhanced couplings to $gg$ and $\gamma\gamma$ due to trace anomaly, can
explain the ratios in Eq.(\ref{exp}).  Also, the radion can give rise to 
enhanced dijet production at $124-126$ GeV.  The associated production
with a $W$ or a $Z$ boson may be observable.  
The radion provides an alternative and the most economical solution to 
explain the observed rates. This is the main result of this work.

A large number of works appeared to interpret the $124-126$ GeV Higgs 
boson in the MSSM framework \cite{mssm}, in the NMSSM framework \cite{nmssm},
in 2HDM \cite{2hdm}, in other SUSY framework \cite{extra}, and others
\cite{others}.

{\it The Radion.}--
The RS model \cite{RS} that uses a warped space-time in a slice of 
extra dimension explains the gauge hierarchy problem well.
The RS model has a four-dimensional massless scalar,
the modulus or radion \cite{GW1,GW2,csaki}, about the background geometry:
\begin{displaymath}
ds^2 = e^{-2 k |\varphi| T(x)} g_{\mu\nu}(x) \, dx^\mu dx^\nu - T^2(x) d\varphi^2 \;,
\end{displaymath}
where $g_{\mu\nu}(x)$ is the four-dimensional graviton and $T(x)$ is the modulus
field.
The most important ingredients of the above brane configuration for phenomenological studies 
are the required size of the modulus field such that it generates the desired weak scale 
from the high scale $M$ and the stabilization of the modulus field at this value.
A stabilization mechanism was proposed by Goldberger and Wise \cite{GW1} 
that a bulk scalar field propagating in the background solution of the
metric can generate a potential that can stabilize the modulus field.  The
minimum of the potential can be arranged to give the desired value of 
$k r_c$ without fine-tuning of parameters.
It has been shown \cite{GW2} that if a large value of $k r_c \sim 12$, needed
to solve the hierarchy problem, arises from a small bulk scalar mass, then the
modulus potential near its minimum is nearly flat for values of the modulus 
vacuum expectation value
that solves the hierarchy problem.  As a consequence, besides getting a mass,
the modulus field 
is likely to be lighter than any Kaluza-Klein modes of any bulk
field.  The lightest mode is the radion, which has a mass of the 
order of 100 GeV 
to a TeV, and the strength
of its coupling to the SM fields is of the order of $O(1/{\rm TeV})$.
There is no theoretical preferred mass region for the radion, and it was
shown in Ref.~\cite{radion-Higgs} that the unmixed radion is consistent 
with electroweak precision data.
Therefore, the detection of this radion may be the first
signature of the RS model.

The interactions of the radion $\phi$ with the SM particles on the brane are 
model-independent and are governed by four-dimensional general covariance 
given by the following Lagrangian:
\begin{equation}
\label{T}
{\cal L}_{\rm int} = \frac{\phi}{\Lambda_\phi} \; T^\mu_\mu ({\rm SM}) \;,
\end{equation}
where $\Lambda_\phi= \langle \phi \rangle$ is of the order of TeV and $T_\mu^\mu$
is the trace of the SM energy-momentum tensor, which is given by
\begin{eqnarray}
T^\mu_\mu ({\rm SM}) &=& \sum_f m_f \bar f f - 2 m_W^2 W_\mu^+ W^{-\mu} 
-m_Z^2 Z_\mu Z^\mu \nonumber \\
 && + (2m_H^2 H^2 - \partial_\mu H \partial^\mu H  ) + \cdots \;,
\end{eqnarray}
where $\cdots$ denotes higher order terms.
The couplings of the radion with fermions $f$, gauge bosons $W$ and $Z$, 
and Higgs boson $H$ are completely fixed by Eq. (\ref{T}). 

For the coupling of the radion to a pair of gluons (photons), there are
contributions from 1-loop diagrams with the top quark (top quark 
and $W$) in the loop as well as from the trace anomaly.
The contribution from the trace anomaly for gauge fields is given by
\begin{equation}
T^\mu_\mu({\rm SM})^{\rm anom} = \sum_a \frac{\beta_a (g_a)}{2g_a} 
F_{\mu\nu}^a F^{a \mu\nu} \;.
\end{equation}
For QCD, $\beta_{\rm QCD}/2g_s = -(\alpha_s/8\pi) b_{\rm QCD}$, where
$b_{\rm QCD} = 11 - 2 n_f/3$ with $n_f=6$.  Thus, 
the effective coupling of $\phi g(p_1) g(p_2)$, including the 1-loop diagrams 
of the top quark and the trace anomaly contributions, is given by
\begin{equation}
\label{rgg}
\frac{i \delta_{ab}\alpha_s}{2\pi \Lambda_\phi}
\left[ b_{\rm QCD} + y_t ( 1+ (1-y_t)f(y_t) ) \right ]
\left( p_1 \cdot p_2 g_{\mu \nu} - p_{2_\mu} p_{1_\nu} \right )
\end{equation}
where $y_t= 4 m_t^2/2p_1 \cdot p_2$ with the gluon incoming momenta $p_1$ and 
$p_2$.
Similarly, 
the effective coupling of $\phi \gamma(p_1)\gamma(p_2)$, 
including the 1-loop diagrams 
of the top quark and $W$ boson and the trace anomaly contributions, 
is given by
\begin{eqnarray}
\label{rgammagamma}
&&\frac{i \alpha_{\rm em}}{2\pi \Lambda_\phi}
\left[  b_2  + b_Y -  (2+3y_W +3y_W (2-y_W)f(y_W) ) \right. \nonumber \\
&& \left. \hspace{-0.2in} + \frac{8}{3} \, y_t ( 1+ (1-y_t)f(y_t) ) \right] 
\times \left( p_1 \cdot p_2 g_{\mu \nu} - p_{2_\mu} p_{1_\nu} \right) \;,
\end{eqnarray}
where 
$b_2=19/6$, $b_Y=-41/6$ and $y_i= 4 m_i^2/2p_1 \cdot p_2$ 
with $i = W,t$.
In the above Eqs.(\ref{rgg}) and (\ref{rgammagamma}), 
the function $f(z)$ is given by
\begin{displaymath}
f(z) = \left \{ \begin{array}{cr}
\left[ \sin^{-1} \left(\frac{1}{\sqrt{z}} \right ) \right ]^2\;, & z \ge 1 \\
-\frac{1}{4} \left[ \log \left( \frac{1+\sqrt{1-z}}{1-\sqrt{1-z}} \right) - i \pi \right ]^2
\;, & z <1 
\end{array}
\right . \;.
\end{displaymath}
There have been many phenomenological studies of the radion 
or dilaton at colliders
\cite{coll} in the literature.  More recent works related to the LHC 
can be found  in Ref.~\cite{more}.

{\it Decays and Production of the Radion.}--
With the above interactions, we can calculate the partial widths of the radion
into $gg$, $\gamma\gamma$, $f\bar f$, $W^+W^-$, $ZZ$, and $HH$. 
The partial widths are given by
\begin{equation}
\Gamma(\phi \to gg) = \frac{\alpha_s^2 m^3_\phi}{32 \pi^3 \Lambda_\phi^2}
 \bigg| b_{\rm QCD} + x_t( 1+ (1-x_t) f(x_t) ) \bigg|^2 , 
\end{equation}
\begin{eqnarray}
&&\hspace{-0.3in}
\Gamma(\phi \to \gamma\gamma) = \frac{\alpha_{\rm em}^2 m^3_\phi}
{256 \pi^3 \Lambda_\phi^2} \Bigg| b_2+b_Y 
  - (2+3 x_W +3x_W  \nonumber\\ 
 &&  \times (2-x_W)f(x_W))  
 +  \frac{8}{3} x_t( 1+ (1-x_t) f(x_t) ) \Bigg|^2 , 
\end{eqnarray}
\begin{equation}
\Gamma(\phi \to f\bar f) = \frac{N_c m_f^2 m_\phi}{8\pi \Lambda^2_\phi}
(1-x_f)^{3/2} , 
\end{equation}
\begin{equation}
\Gamma(\phi \to W^+ W^-) = \frac{m_\phi^3}{16 \pi \Lambda_\phi^2}
 \sqrt{1- x_W} \left( 1 - x_W +\frac{3}{4} x_W^2 \right ) ,  
\end{equation}
\begin{equation}
\Gamma(\phi \to ZZ) = \frac{m_\phi^3}{32 \pi \Lambda_\phi^2}\;
 \sqrt{1- x_Z} \left( 1 - x_Z +\frac{3}{4} x_Z^2 \right ) ,  
\end{equation}
\begin{equation}
\Gamma(\phi \to HH) = \frac{m_\phi^3}{32\pi \Lambda_\phi^2}\; 
  \sqrt{1-x_H} \left( 1+ \frac{x_H}{2} \right )^2 ,
\end{equation}
where $x_i = 4 m_i^2/m_\phi^2$ $(i=f,W,Z,H)$ and 
$N_c=3 \, (1)$ for quarks (leptons).
Note that the branching ratios are independent of $\Lambda_\phi$.

In calculating the partial widths into fermions, we have used the 3-loop
running masses with scale $Q^2 = m_\phi^2$.   
We have also allowed the off-shell decays of the $W$ and $Z$ bosons and 
that of the top quark. The features of radion decay branching ratios are
similar to the decay of the Higgs boson, except the following. 
At $m_\phi \alt 140$ GeV, the decay width is dominated by $\phi \to gg$, while 
the decay width of the SM Higgs boson is dominated by the 
$b\bar b$ mode.  At larger $m_\phi$, $\phi$ also decays into a pair of
Higgs bosons ($\phi \to HH$) if kinematically allowed, while the SM
Higgs boson cannot.  Similar to the SM Higgs boson, 
as $m_\phi$ goes beyond the $WW$ and $ZZ$ 
thresholds, the $WW$ and $ZZ$ modes dominate with the $WW$ partial width 
about a factor of 2 of the $ZZ$ partial width.  
We list the relevant branching ratios of the radion in Table~\ref{tab1}
for $m_\phi = 123 - 126$ GeV. Just for comparison with the SM Higgs boson,
we also list the branching ratios and production cross sections of the
SM Higgs boson in Table~\ref{tab2} (from Ref.~\cite{higgs-X}).

\begin{figure}[t!]
\centering
\includegraphics[width=3.5in]{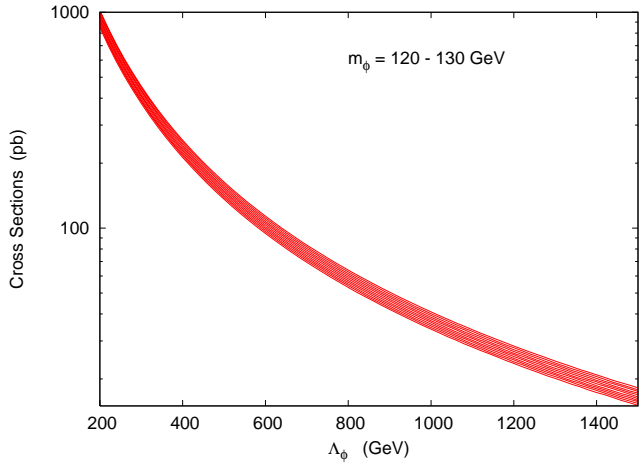}
\caption{\small \label{fig} Production cross sections for $pp \to \phi$
via gluon fusion versus $\Lambda_\phi$ for $m_\phi = 120 - 130$ GeV. 
The top of the ``thick'' curve is for 120 GeV, while the bottom is for 130 GeV.}
\end{figure}

The production channels of the radion at hadronic colliders include
\begin{eqnarray}
gg &\to& \phi \nonumber\\
q \bar q' &\to& W \phi \nonumber \\
q \bar q &\to& Z \phi \nonumber\\
q q' &\to& q q' \phi \;\; \mbox{($WW,ZZ$ fusion)} \nonumber\\
q \bar q \, , \, gg &\to& t \bar t \phi \nonumber \;.
\end{eqnarray}
Similar to the SM Higgs boson, the most important production channel for the
radion is $gg$ fusion.
In addition, $gg\to\phi$ gets further enhancement from the trace anomaly.
We shall consider only the gluon fusion in the following.
We show the production cross sections
for $m_\phi = 120-130$ GeV versus $\Lambda_\phi$ in Fig.~\ref{fig}.

\begin{table*}[t!]
\caption{\small \label{tab1}
The branching ratios of the RS radion 
for $m_\phi = 123 - 126$ GeV. }
\begin{ruledtabular}
\begin{tabular}{c|cccccc}
$m_\phi$  & \multicolumn{6}{c}{Branching ratios} \\
(GeV)  & $gg$ & $b\bar b$ & $\tau\tau$ & $W W^*$ & $ZZ^*$ & $\gamma \gamma$ \\
\hline
 $123$ &  $0.899$ & $0.0608$ & $9.49\times 10^{-3}$ 
       & $0.0246$  & $2.93\times 10^{-3}$  & $0.912 \times 10^{-3}$ \\
 $124$ &  $0.897$ & $0.0598$ & $9.34\times 10^{-3}$ 
       &  $0.0267$  & $3.25\times 10^{-3}$    & $0.918 \times 10^{-3}$ \\
 $125$ &  $0.896$ & $0.0588$ & $9.2\times 10^{-3}$ 
       & $0.0291$  & $3.6\times 10^{-3}$  & $0.925 \times 10^{-3}$ \\
 $126$ &  $0.894$ & $0.0578$ & $9.05\times 10^{-3}$ 
       & $0.0317$  & $3.98\times 10^{-3}$ & $0.931 \times 10^{-3}$ 
\end{tabular}
\end{ruledtabular}
\end{table*}

\begin{table*}[t!]
\caption{\small \label{tab2}
A few production cross sections and branching ratios
of the SM Higgs boson for $m_H = 123 - 126$ GeV. We borrow the values 
from Ref.~\cite{higgs-X}.}
\begin{ruledtabular}
\begin{tabular}{c|ccc|ccccc}
$m_H$  & \multicolumn{3}{c|}{Cross sections (pb)} & 
              \multicolumn{5}{c}{Branching ratios} \\
(GeV) & $gg\to H$ & $WH$ & $ZH$  &  $b\bar b$ & $\tau\tau$ & $W W^*$ & $ZZ^*$ 
    & $\gamma \gamma$ \\
\hline
 $123$ &  $15.8$ & $0.61$ & $0.33$ & $0.607$ & $0.067$
             & $0.185$ & $0.022$  & $2.28\times 10^{-3}$ \\
 $124$ &  $15.6$ & $0.59$ & $0.32$ & $0.592$ & $0.065$
             &$0.200$ & $0.0242$ & $2.29\times 10^{-3}$ \\
 $125$ &  $15.3$ & $0.57$ & $0.32$ & $0.577$ & $0.064$
             &$0.216$ & $0.0266$ & $2.29\times 10^{-3}$ \\
 $126$ &  $15.1$ & $0.56$ & $0.31$ & $0.561$ & $0.062$
           & $0.233$ & $0.0291$ & $2.29\times 10^{-3}$ 
\end{tabular}
\end{ruledtabular}
\end{table*}

{\it Comparison to the LHC data.}--
Using the appropriate entries from Tables \ref{tab1} and \ref{tab2} 
for $m_{\phi / H} = 124$ GeV, 
we can compute the following ratio:
\begin{equation}
\frac{\sigma(\phi) \times B(\phi \to \gamma\gamma) }
     {\sigma(H) \times B(H \to \gamma\gamma) } =
\frac{\sigma(\phi) \times 0.918\times 10^{-3} }
     {15.6 \,{\rm pb} \times 2.29\times 10^{-3} }  = 2.1
\end{equation}
where 2.1 is the central value of the CMS 
data \cite{cms} for this ratio. 
Therefore, the value of $\sigma(\phi)$ obtained in the above equation is 
82 pb, which corresponds to $\Lambda_\phi =0.68$ TeV from 
Fig. \ref{fig}.  
If we use the diphoton data with error bars $(2.1\,^{+\,0.6}_{-\,0.7})$,
the corresponding $\Lambda = 0.68 \,^{+\,0.15}_{-\,0.08}$ TeV.
Once the ratio for 
$\gamma\gamma$ is fixed, the other ratios can be easily 
obtained by using Tables \ref{tab1} and \ref{tab2}:
\begin{eqnarray}
&&\frac{\sigma(\phi)  B(\phi \to b\bar b)}
     {\sigma(H)  B(H \to b\bar b)} =
\frac{82 \,{\rm pb} \times 0.0598 }
     {15.6 \,{\rm pb} \times  0.592} = 0.53,  \\
&& \hspace{-0.2in} \frac{\sigma(\phi)  B(\phi \to \tau\tau)}
     {\sigma(H)  B(H \to \tau\tau)} =
\frac{82 \,{\rm pb} \times 9.34 \times 10^{-3} }
     {15.6 \,{\rm pb} \times 0.065  } = 0.75,  \\
&& \hspace{-0.2in} \frac{\sigma(\phi)  B(\phi \to WW^*)}
     {\sigma(H) B(H \to WW^*)} =
\frac{82 \,{\rm pb} \times 0.0267 }
     {15.6 \,{\rm pb} \times  0.200} = 0.70,  \\
&& \hspace{-0.3in} \frac{\sigma(\phi)  B(\phi \to ZZ^*)}
     {\sigma(H) B(H \to ZZ^*)} =
\frac{82 \,{\rm pb} \times 3.25 \times 10^{-3} }
     {15.6 \,{\rm pb} \times 0.0242} = 0.70.  
\end{eqnarray}
Therefore, we obtain a set of ratios
which match well with the central values of 
the CMS data, provided that we first match the 
$\gamma\gamma$ mode to the experimental data.  We repeat the exercise
for other $m_\phi = 123-126$ GeV with the results shown in Table~\ref{tab3}. 
The ratios vary very little in this radion mass range.

\begin{table}[t!]
\caption{\small \label{tab3}
The ratio 
$\frac{\sigma(\phi) \times B(\phi \to X)}
     {\sigma(H) \times B(H \to X)}$ for $m_{\phi / H} = 123 -126$ GeV.
}
\begin{ruledtabular}
\begin{tabular}{c|ccccc}
$m_{\phi / H}$  &  \multicolumn{4}{c}{$\frac{\sigma(\phi) \times B(\phi \to X)}
     {\sigma(H) \times B(H \to X)}$} \\
(GeV) & $\gamma \gamma$  &  $b\bar b$ & $\tau\tau$ & $W W^*$ & $ZZ^*$ \\ 
\hline
 $123$ &  $2.1$ & $0.53$ & $0.74$ & $0.70$ & $0.70$\\
 $124$ &  $2.1$ & $0.53$ & $0.75$ & $0.70$ & $0.70$\\
 $125$ &  $2.1$ & $0.53$ & $0.75$ & $0.70$ & $0.70$\\
 $126$ &  $2.1$ & $0.53$ & $0.75$ & $0.70$ & $0.71$
 \end{tabular}
\end{ruledtabular}
\end{table}

\begin{table}[t!]
\caption{\small \label{tab4}
Production cross sections in femtobarns for $\sigma (W\phi \to W gg)$ and 
$\sigma (Z\phi \to Z gg)$ at the Tevatron and at the LHC-7. 
The $\Lambda_\phi$ is set at 0.68 TeV.}
\begin{ruledtabular}
\begin{tabular}{c|cc|cc}
$m_{\phi}$  &  \multicolumn{2}{c|}{$\sigma(W\phi) \times B(\phi \to gg)$} &
              \multicolumn{2}{c}{$\sigma(Z\phi) \times B(\phi \to gg)$}  \\
(GeV) &  Tevatron & LHC-7  & Tevatron & LHC-7 \\
\hline
 $123$ &  $19.1$ & $73.4$ & $11.5$ & $39.1$\\
 $124$ &  $18.5$ & $71.2$ & $11.1$ & $38.0$\\
 $125$ &  $17.9$ & $69.2$ & $10.8$ & $36.9$\\
 $126$ &  $17.4$ & $67.1$ & $10.5$ & $35.8$
 \end{tabular}
\end{ruledtabular}
\end{table}

\bigskip

{\it Implications.}--
The radion has a large branching ratio into $gg$, which will give rise to
a dijet signal at the Tevatron and the LHC. The cross section 
$\sigma(gg\to \phi)\times B(\phi \to gg) \approx 73$ pb 
at the LHC and only 3.4 pb at the Tevatron.  The huge QCD background 
will overwhelm the dijet signal.  The only possibility is to consider
the associated production with a $W$ or a $Z$ boson.  
In Table~\ref{tab4}, we calculate the
production cross section of $W\phi$ and $Z\phi$ at the Tevatron 
and at the LHC, multiplied by the gluonic branching ratio.
With this level of cross sections it is still difficult to beat 
the $Wjj$ and $Zjj$ background. If the systematics can be reduced to 
a few percent level, it may have some chance to see this dijet signal.

In conclusion, the CMS and ATLAS Collaborations 
have seen excess in a number of channels;
in particular, the $\gamma\gamma$ channel 
has a cross section about twice the 
SM value, while the other channels are slightly suppressed (by
about $0.5-0.7$) relative to their SM values. We have proposed
the RS radion as a possible candidate for the particle observed.  
While it is not easy to accommodate a 125 GeV Higgs boson with 
an enhanced diphoton rate in the MSSM \cite{mssm}, NMSSM \cite{nmssm},
and a number of other popular models \cite{little,inert,see-saw},
the RS radion provides the most economical way to interpret the data.
In order to give a $124 - 126$ GeV Higgs boson within MSSM, the stop
sector must consist of a large mixing that gives rise to  
one very heavy stop $\tilde t_2$ and one relatively light stop $\tilde t_1$. 
Within the MSSM and NMSSM, it is rather difficult
to enhance the $\gamma\gamma$ production rate \cite{mssm,nmssm}; only
in some less restrictive NMSSM can the rate be enhanced by a factor 
up to 2 \cite{nmssm}.
The littlest Higgs model \cite{little} always gives a slight reduction in
the diphoton rate. The inert Higgs doublet model \cite{inert} gives
a diphoton rate in the range of $0.8 - 1.3$ relative to the SM rate, but 
it can hardly go over $1.5$.   In a type-II seesaw model \cite{see-saw},
the diphoton production rate can be enhanced significantly because of the
contribution from the double-charged Higgs boson 
but at relatively large values of self-couplings.
The radion, due to the trace anomaly,
has enhanced couplings to a pair of photons and gluons. Thus, the production
rate of $\sigma(\phi) \times B(\phi \to \gamma\gamma)$ can be enhanced 
relative to the SM cross section.  The data requires $\Lambda_\phi \approx 
0.68$ TeV.  At the same time, the other channels $b\bar b$, $\tau\tau$,
$WW$, and $ZZ$ are all suppressed by a factor of $0.5 - 0.7$ (shown in
Table~\ref{tab3}) relative to the SM.  Therefore, the RS radion provides a 
reasonably good interpretation to the data.  Such a radion will give 
rise to a large dijet resonance signal, though it is still very difficult
to identify it in the presence of huge QCD background, unless
the systematic uncertainty can be reduced to a few percent level.  

This work was supported in part by the National Science Council of
Taiwan under Grants No. 99-2112-M-007-005-MY3 and No.
98-2112-M-001-014-MY3 as well as the
WCU program through the KOSEF funded by the MEST (R31-2008-000-10057-0).


\end{document}